\documentclass{emulateapj}
\usepackage{apjfonts}
\usepackage{psfig}
\usepackage{mathrsfs}
\usepackage{amsmath}

\begin{document}

\slugcomment{AJ, in press}
\shorttitle{NGC~5286. I. CMD}
\shortauthors{M. Zorotovic et al.}

\title{The Globular Cluster NGC~5286.\\I. A New CCD {\it{BV}} Color-Magnitude Diagram\footnote{Based on observations obtained with the 1.3m Warsaw telescope at the Las Campanas Observatory, Chile.}}

\author{M. Zorotovic,\altaffilmark{1} M. Catelan,\altaffilmark{1} M. Zoccali\altaffilmark{1}
  B. J. Pritzl,\altaffilmark{2} H. A. Smith,\altaffilmark{3}\\ A. W. Stephens,\altaffilmark{4}
  R. Contreras,\altaffilmark{1,5} M. E. Escobar\altaffilmark{1}
}

\altaffiltext{1}{Pontificia Universidad Cat\'olica de Chile, Departamento de 
       Astronom\'\i a y Astrof\'\i sica, Av. Vicu\~{n}a Mackenna 4860, 
       782-0436 Macul, Santiago, Chile; e-mail: mzorotov, mcatelan, mzoccali, mescobar@astro.puc.cl}

\altaffiltext{2}{Department of Physics and Astronomy, University of Wisconsin Oshkosh, Oshkosh, WI 54901;  
e-mail: pritzlb@uwosh.edu}

\altaffiltext{3}{Department of Physics and Astronomy, Michigan State University, East Lansing, MI 48824;  
e-mail: smith@pa.msu.edu}

\altaffiltext{4}{Gemini Observatory, 670 N. A\'ohoku Place, Hilo, HI 96720; e-mail: astephens@gemini.edu}

\altaffiltext{5}{Current address: INAF, Osservatorio Astronomico di Bologna, via Ranzani 1, I-40127 Bologna, 
Italy; e-mail: rodrigo.contreras@oabo.inaf.it}

\begin{abstract}
We present {\it{BV}} photometry of the Galactic globular cluster \objectname{NGC~5286}, based on 128 $V$ frames and 133 $B$ frames, and covering the entire face of the cluster. Our photometry reaches almost two magnitudes below the turn-off level, and is accordingly suitable for an age analysis. Field stars were removed statistically from the cluster's  color-magnitude diagram (CMD), and a differential reddening correction applied, thus allowing a precise ridgeline to be calculated.  

Using the latter, a metallicity of ${\rm [Fe/H]} = -1.70 \pm 0.10$ in the \citeauthor{zw1984} scale, and ${\rm [Fe/H]} = -1.47\pm 0.02$ in the \citeauthor{CG1997} scale, was derived on the basis of several parameters measured from the red giant branch, in good agreement with the value provided in the \citeauthor{harr1996} catalog. 

Comparing the \objectname{NGC~5286} CMD with the latest photometry for \objectname{M3} by P. B. Stetson (2008, priv. comm.), and using VandenBerg isochrones for a suitable chemical composition, we find evidence that \objectname{NGC~5286} is around $1.7 \pm 0.9$~Gyr older than \objectname{M3}. This goes in the right sense to help account for the blue horizontal branch of NGC~5286, for which we provide a measurement of several morphological indicators. If NGC~5286 is a bona fide member of the Canis Major dwarf spheroidal galaxy, as previously suggested, our results imply that the latter's oldest components may be at least as old as the oldest Milky Way globular clusters. 
 
\end{abstract}

\keywords{stars: Hertzsprung-Russell diagram --- stars: variables: other --- Galaxy: globular clusters: individual (NGC~5286, NGC~5272) --- galaxies: dwarf --- galaxies: star clusters}

\section{Introduction}\label{sec:intro}

Globular clusters (GCs) are among the very oldest objects in the Universe. Accordingly, detailed studies of their stellar contents, including both the variable and non-variable components, hold the key to the formation and early evolution of galaxies~-- and of our own Galaxy in particular. RR Lyrae variable stars, which are present in GCs in large numbers \citep[e.g.,][]{ccea01}, play a particularly important role in determining the extent to which the Galaxy  may have formed from the accretion of smaller ``protogalactic fragments,'' as currently favored by $\Lambda$CDM cosmology \citep[e.g.,][]{mc05,mc07}. However, while over 150 Galactic GCs are currently known \citep[see, e.g.,][and references therein]{jbea07}, only a relatively small fraction has been surveyed for stellar variability using state-of-the-art techniques, including CCD detectors and image subtraction algorithms. We have accordingly started an extensive variability survey of GCs \citep[e.g.,][]{mcea06}, with the ultimate goal to help put constraints on the way the Galaxy has formed. As a natural by-product of our variability searches, deep and high-precision color-magnitude diagrams (CMDs) are obtained, often covering large areas over the face of the observed clusters. This affords a new look into the clusters' physical parameters, as implied by these CMDs.

\objectname{NGC~5286} (C1343-511) is a particularly interesting object in this context. This is a bright ($M_V = -8.26$) and fairly compact GC (with a core radius of $0.29'$, a half-light radius of only $0.69'$, and a tidal radius of $8.36'$), of intermediate metallicity (${\rm [Fe/H]} = -1.67$) in Centaurus (all values from \citealp{harr1996}). It may be associated with the Canis Major dwarf spheroidal galaxy (\citeauthor{pfea04} \citeyear{pfea04}; see also \citeauthor*{dfea04} \citeyear{dfea04}). Unfortunately, \objectname{NGC~5286} has received relatively little attention, the latest studies to produce a CMD having been the one by \citet{nsea95}, which relied on a CCD with $512 \times 320$ pixels$^2$, covering a field of view of only $3.1 \times 1.9 \, {\rm arcmin}^2$, and avoiding the cluster center; and the one by \citet{ebea96}, which~-- though covering a larger area over the face of the cluster~-- did not produce a sufficiently deep CMD as to provide a reliable ridgeline extending below the main-sequence (MS) turnoff (TO) point (see their Fig.~8). One of the reasons for the relative neglect of this cluster may have been its relatively high foreground reddening, $E(\bv) = 0.24$ \citep{harr1996}. In the same vein, \objectname{NGC~5286} has never been properly studied, using state-of-the-art techniques, in terms of its variable star content, even though it is known to harbor over a dozen RR Lyrae variables \citep[e.g.,][]{ccea01}.  

The main purpose of the present study is to provide the first extensive, CCD-based variability study of \objectname{NGC~5286}, which leads to a deep CMD covering a much larger field over the face of the cluster than has  been available thus far. In the present paper we shall report on the CMD and the derived cluster parameters, whereas a companion paper (Zorotovic et al. 2008, hereafter Paper II) will describe in detail the variability search and derived variable star parameters for the 56 variables that were found in \objectname{NGC~5286} (which increases by a factor of $\approx 3$ the number of known/measured variables in the cluster). 

We begin in \S\ref{sec:obs} by describing the acquired data and reduction techniques. In \S\ref{sec:CMD} we describe our derived CMD, along with the physical parameters of the cluster obtained therefrom. \S\ref{sec:vsM3} is devoted to a differential age comparison with \objectname{M3}, a cluster of similar metallicity to \objectname{NGC~5286}. We finally close in \S\ref{sec:summ} by summarizing the results of our investigation. 

\section{Observations and Data Reduction} \label{sec:obs}
Time-series $B$, $V$ images of \objectname{NGC~5286} were collected over a one-week run in April 2003, with the $1.3$m Warsaw University Telescope at Las Campanas Observatory (LCO) and using the ``second generation'' CCD mosaic camera, commisioned in May 2001. The 8kMOSAIC camera consists of eight thin SITe $2048\times 4096$ CCD chips ($8192\times 8192$ pixels of $0.26$~arcsec/pixel), giving a total field of view equal to $35' \times 35'$. The readout time of the camera is 98 seconds, with readout noise of 6 to 9~$e^-$ (depending on the chip) and gain of 1.3 $e^-$/ADU. A total of 128 frames in {\it{V}} and 133 frames in {\it{B}} were thus acquired. 
 
In this work we present the results for chip 2, covering a field of $9' \times 17'$, where the cluster dominates. We use chip 3, covering a similar field, to select field stars to perform statistical decontamination of the cluster.

Photometry was performed using DAOPHOT II/ALLFRAME \citep{Stet1987,pbs94}. Our calibration is the same as presented by \citet{Cetal2005} and Contreras et al. (2008, in preparation) in their study of M62 (NGC~6266), and by Escobar et al. (2008, in preparation) in their study of M69 (NGC~6637). These two clusters were observed during the same nights, and with the same telescope/instrument combination, as NGC~5286. To obtain the employed calibration, several different \citet{al92} fields were observed during photometric nights, including PG+0918, PG+1323, PG+1525, PG+1528, PG+1633, PG+1657, and RU~152. These fields were centered on chip 2, where most of the stars in the observed clusters were found. A total of 117 and 157 Landolt standard stars in $B$ and $V$, respectively, were used to derive our final calibrations. The derived equations for $B$ and $V$ had root-mean-square deviations of order 0.18~mag in $B$ and 0.04~mag in $V$. The derived calibration was then propagated to chip 3, by using about 660 stars in common between chips 2 and 3. Further details of the derived calibrations will be provided in Contreras et al. (2008) and Escobar et al. (2008). 


\section{Color-Magnitude Diagram} \label{sec:CMD}
\subsection{Field-Star Decontamination}

\begin{figure}[t]                            
      \plotone{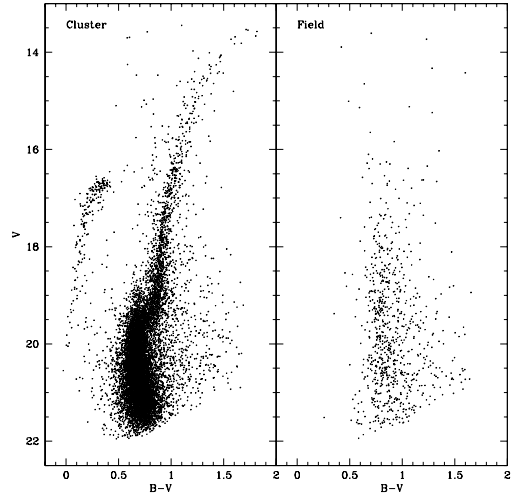}
      \caption{\footnotesize{{\em  Left}:  ($\bv$,{\it{V}})  CMD   for
      NGC~5286   before  field-star   decontamination.  The  innermost
      cluster regions ($r  \leq  0.3'$)  are not shown.  {\em  Right}:
      ($\bv$,{\it{V}}) CMD  for field  stars   used to carry  out  the
      decontamination.}}  \label{rawCMD} \end{figure}

Our derived  CMD for the main  cluster field  (chip 2), after removing
the innermost cluster regions ($r \leq 0.3'$) which are badly affected
by    crowding,   is  shown     in  Figure~\ref{rawCMD}    ({\em left
panel}). Known variables (see Paper II) were removed from this CMD. As
can be seen, field contamination is not negligible, which may adversly
affect   our determination  of  the cluster   ridgeline, and thus  the
measurement of key physical parameters for  the cluster (including its
age and  metallicity).  We have  accordingly resorted to  performing a
statistical  decontamination of the  observed   CMD, using the  method
described in \citet{gallart03}, with a maximum distance $d=0.15$ mag.

	  \begin{figure}[ht]
      \plotone{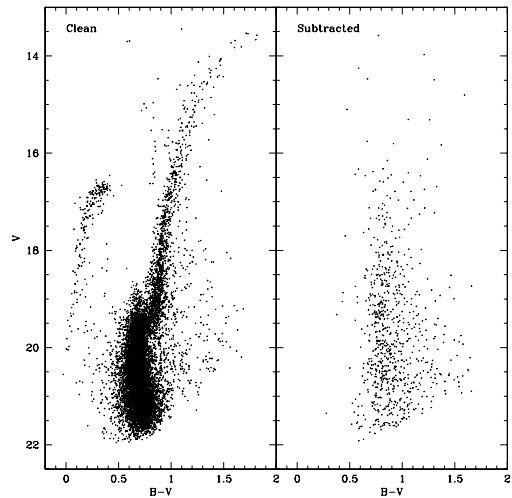}
      \caption{\footnotesize{{\em Left:} The CMD of NGC~5286, after
       statistical decontamination from field stars. {\em Right:}
       Stars subtracted from the original CMD (Fig.~\ref{rawCMD}, left)
       in order to obtain the decontaminated CMD shown on the left.}}
      \label{decontCMD}
      \end{figure}

\begin{figure}[t]
      \includegraphics[angle=-90, scale=0.34]{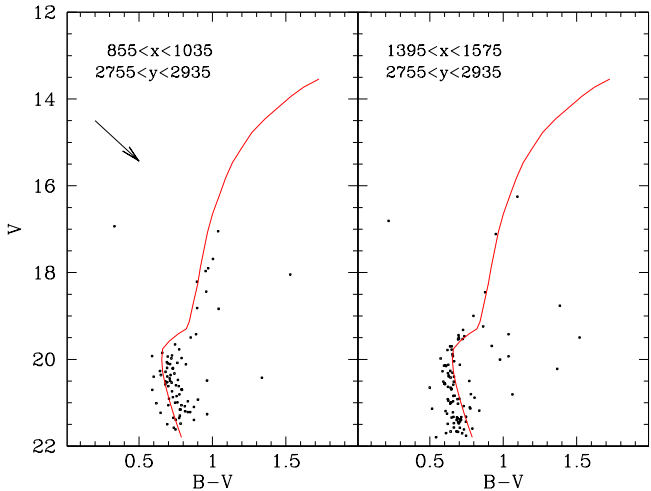}
      \epsscale{1.0}
      \caption{\footnotesize{Examples of CMDs for two small 
      areas located in different regions across the face of the cluster ({\em see
      labels}). Clearly, the two CMDs are shifted along the reddening vector (shown 
	  by the arrow) with respect to the reference mean locus, indicating the presence
	  of differential reddening across the face of the cluster.}
	  }
      \label{redd}
      \end{figure}

We use as control field the stars  from chip 3   that lie outside the
tidal radius; their corresponding  CMD is shown in Figure~\ref{rawCMD}
({\em   right panel}).   Figure~\ref{decontCMD}   shows the  resulting
($\bv$,{\it{V}}) CMD, after decontamination. As  can be seen from
this figure, the main  branches still reveal  some spread, which  does
not decrease substantially  as one goes towards brighter magnitudes~--
as would   have been expected if it   were due to   photometric errors
alone.   We therefore suspect  that  a small amount of differential
reddening may  be  the culprit. To   check  this, we have  mapped  the
cluster  CMD   following   the   procedure   described,    e.g.,    in
\citet{gpea99}. Indeed, Figure~\ref{redd} shows
that different regions of chip~3 yield CMDs that are shifted, with 
respect to a reference mean locus, along the reddening vector. It does 
appear, therefore, that differential reddening is present in NGC~5286, 
at the level of a few hundredths of a magnitude in $E(\bv)$.

\begin{figure}[t]
      \includegraphics[angle=-90, scale=0.34]{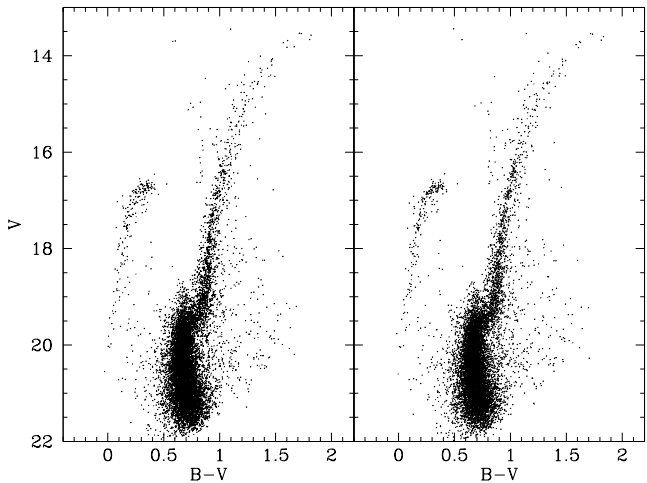}
      \caption{\footnotesize{Comparison between the original,
      decontaminated cluster CMD ({\em left}) and the one corrected for differential
      extinction across the cluster area ({\em right}).}
	  }
      \label{reddveri}
      \end{figure}

To produce a tighter CMD, a correction was applied to minimize this effect, 
following the procedure described in \citet{gpea99}. The resulting CMD is 
shown in Figure~\ref{reddveri}. Due to the saturation of points in this 
plot, the difference is not so striking in the MS, while the red giant 
branch (RGB) and the blue horizontal branch (HB) do appear significantly
narrower.

\begin{figure}[t]
      \includegraphics[angle=-90, scale=0.34]{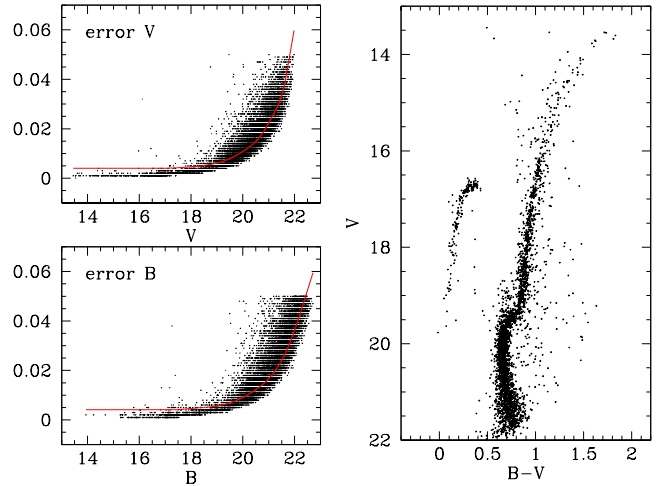}
      \caption{\footnotesize{Selection based on the photometric errors ({\em left}), 
	  and final CMD for NGC~5286 ({\em right}).}
         }
      \label{err}
      \end{figure}

Finally, in order to further improve the CMD, stars with too large
errors in both $B$ and $V$ were excluded. As shown in 
Figure~\ref{err}, the applied error cut follows the shape of the lower 
envelope of the error distribution (Poisson error), thus excluding
the stars with large errors {\it compared} to the error expected at 
their respective magnitude levels. The resulting CMD is shown in the 
right panel of Figure~\ref{err}. This is our final CMD for the cluster, 
and the one used in the subsequent analysis.

Before proceding, we note that 
our CMD  is  clearly  in   good qualitative  agreement with  the  ones
previously published by \citet{nsea95} and \citet{ebea96}, revealing a
predominantly blue  HB and  a moderately steep RGB indicative of an 
intermediate-to-low metallicity.
      
\begin{figure}[t]
      \plotone{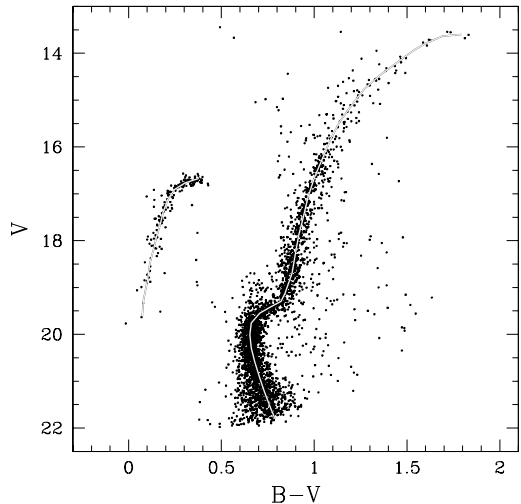}
      \caption{
($\bv$,{\it{V}}) CMD for NGC~5286, with the derived ridgelines overplotted.}
      \label{ridgeline}
      \end{figure}

To determine the mean ridgelines, we have experimented 
with several different techniques, but finally decided upon what essentially amounts 
to an eye fit to the data, similar to the procedure previously followed, for instance, 
by \citet{lbea00} and \citet*{psea05}. Table~\ref{tablaridge} presents the adopted normal points for each 
branch. In Figure~\ref{ridgeline} the ($\bv$,{\it{V}}) CMD is shown with the mean 
ridgelines overplotted.

To determine the position of the turn off we fit a parabola to a small region of the MS near the TO point. The MS TO point is found to be at $V_{\rm TO} = 20.05 \pm 0.1$ mag and $(\bv) = 0.66 \pm 0.02$.

\begin{table}[t]
\begin{center}
\footnotesize
\caption{\footnotesize{Mean Fiducial Points for NGC~5286}}
\begin{tabular}{lc}
\tableline\tableline
{\it{V}} & (\bv) \\
\tableline
\tableline
\multicolumn{2}{c}{MS + SGB\tablenotemark{a} + RGB}\\
\tableline
$13.602$........................... & $1.793$ \\
$13.636$........................... & $1.692$ \\
$13.785$........................... & $1.602$ \\
$13.933$........................... & $1.537$ \\
$14.213$........................... & $1.441$ \\
$14.464$........................... & $1.356$ \\
$14.771$........................... & $1.268$ \\
$15.135$........................... & $1.197$ \\
$15.470$........................... & $1.136$ \\
$15.819$........................... & $1.090$ \\
$16.210$........................... & $1.048$ \\
$16.643$........................... & $1.002$ \\
$17.076$........................... & $0.966$ \\
$17.495$........................... & $0.941$ \\
$17.886$........................... & $0.917$ \\
$18.263$........................... & $0.899$ \\
$18.696$........................... & $0.880$ \\
$19.129$........................... & $0.842$ \\
$19.297$........................... & $0.821$ \\
$19.409$........................... & $0.768$ \\
$19.546$........................... & $0.707$ \\
$19.758$........................... & $0.661$ \\
$20.009$........................... & $0.654$ \\
$20.288$........................... & $0.661$ \\
$20.582$........................... & $0.679$ \\
$20.903$........................... & $0.704$ \\
$21.224$........................... & $0.732$ \\
$21.560$........................... & $0.764$ \\
$21.797$........................... & $0.789$ \\
\tableline
\multicolumn{2}{c}{Blue HB}\\
\tableline
$16.671$........................... & $ 0.378$  \\  
$16.769$........................... & $ 0.303$  \\  
$16.908$........................... & $ 0.247$  \\  
$17.118$........................... & $ 0.211$  \\  
$17.383$........................... & $ 0.186$  \\  
$17.635$........................... & $ 0.172$  \\  
$17.914$........................... & $ 0.150$  \\  
$18.207$........................... & $ 0.133$  \\  
$18.487$........................... & $ 0.115$  \\  
$18.864$........................... & $ 0.101$  \\  
$19.283$........................... & $ 0.080$  \\  
$19.618$........................... & $ 0.073$  \\
\tableline
\tablenotetext{a}{ SGB = subgiant branch.} 

\end{tabular}
\label{tablaridge}
\end{center}
\end{table}       

\subsection{Metallicity and Reddening}\label{sec:metred}
 \citet[][hereafter F99]{Fer1999}  derived a set of metallicity indicators in terms of RGB parameters. We use equations from Table~4 in F99 to estimate the metallicity for \objectname{NGC~5286} in the \citet[][hereafter CG97]{CG1997}
scale, ${\rm [Fe/H]}_{\rm CG97}$, and the global metallicity, ${\rm [M/H]}$. The latter takes into due account the effects of an enhancement in the abundance of the so-called $\alpha$-capture elements with respect to the solar proportions \citep*[e.g.,][]{msea93}.

Since several of these indicators require a precise definition of the HB level, before proceeding we must first determine the HB level in the same way as in F99. These authors adopt for the HB the zero-age HB (ZAHB) level in their measurements. The latter is defined as the magnitude of the lower envelope of the observed HB distribution in the region with $0.2 < (\bv)_0 < 0.6$. As we do not have many non-variable stars in this region, we decided to use a theoretical ZAHB 
\citep*[from][for a chemical composition ${\rm [Fe/H]} = -1.71$, ${\rm [\alpha/Fe]} = 0.3$]{VBD2006} 
to match the lower envelope of the HB distribution. We use a reddening of $E(\bv) = 0.24$ \citep{harr1996} and we vertically shifted the simulated ZAHB to match the lower envelope of the observed HB, arriving at the best match that is shown in Figure~\ref{finalhb}. The average ZAHB magnitude over the quoted color range provides us with our final $V_{\rm ZAHB} = 16.69$.

For \objectname{NGC~5286}, we used the mean ridgeline to measure the following RGB parameters: ${\it{(B-V)}}_{0,g}$ (RGB color at the HB level); $\Delta V_{1.1}$, $\Delta V_{1.2}$, and $\Delta V_{1.4}$ [magnitude difference between the HB and RGB at fixed colors ${\it{(B-V)}}_0 = 1.1$, $1.2$, and $1.4$, respectively];	and the two RGB slopes $S_{2.5}$ and $S_{2.0}$ (slope of the line connecting the intersection of the RGB and HB with the points along the RGB located 2.0 and 2.5 mag brighter than the HB, respectively). 

To infer the metallicity from these indices, colors had first to be corrected by extinction. In the Feb.\ 2003 edition of the \citet{harr1996} catalog, a value ${\it{E(\bv)}} = 0.24$ is provided. For comparison, on the basis of the \citet*{dsea98} dust maps one finds $E(\bv) = 0.292$~mag. However, for high reddening values there is some indication that the \citeauthor{dsea98} maps may overestimate the reddening  value. Using equation~(1) in \citet*{pbea00} to correct for this, one gets a corrected reddening value of $E(\bv) =  0.225$~mag. While \citet{Z1985} gives an $E(\bv) =  0.27$~mag for the cluster, \citet{rw85} gives instead $E(\bv) =  0.21$~mag. \citet*{brea88}, on the other hand, list two possible reddening values for the cluster, namely $E(\bv) =  0.28$~mag and $E(\bv) =  0.24$~mag. Taking a straight average over all the quoted values (except the high original value from the \citeauthor{dsea98} maps), one finds for NGC~5286 a suggested average value of $E(\bv) \simeq 0.24 \pm 0.03$. We shall accordingly assume $E(\bv) = 0.24$ for NGC~5286. 

The individual metallicity values derived from each of the indices defined above are listed in Table~\ref{met}.\footnote{Note that equation~(4.6) in F99 actually calibrates $S_{2.0}$ in terms of ${\rm [M/H]}$, whereas their equation~(4.13) calibrates the same quantity in terms of ${\rm [Fe/H]}$ ~ -- and not the other way around, as incorrectly stated in their paper. We thank F.\ Ferraro for his help in clarifying this point.} A median over the listed results gives for NGC~5286 a metallicity ${\rm [Fe/H]}_{\rm CG97} = -1.47 \pm 0.02$ and ${\rm [M/H]} = -1.26 \pm 0.02$ (standard deviation of the mean). That give us a value of ${\rm [Fe/H]} = -1.70 \pm 0.10$ in the \citet[][hereafter ZW84]{zw1984} scale, in good agreement with \citet{harr1996}.

As a check of the adopted reddening and metallicity values, we have also applied the simultaneous reddening and metallicity (SRM) method of \citet{saraj94}, as ported to the $V$, $\bv$ plane by F99. The slopes $S_{2.5}$ and $S_{2.0}$ do not depend on the reddening. Using the average values of ${\rm [Fe/H]}$ and ${\rm [M/H]}$ based only on these two slopes, we obtain $(B-V)_{0,g} = 0.761$ and $(B-V)_{0,g} = 0.765$, respectively~-- in excellent agreement with the values found in Table~\ref{met}. We conclude that the assumed reddening value of $E(\bv) =  0.24$~mag must be close to the correct value for the cluster.

\begin{table}[t]
\begin{center}
\footnotesize
\caption{\footnotesize{NGC~5286 Metallicity}}
\begin{tabular}{lcc}
\tableline\tableline
Parameter & ${\rm [Fe/H]_{CG97}}$ & ${\rm [M/H]}$] \\
\tableline
$(B-V)_{0,g} = 0.762$    & $-1.48$ & $-1.27$ \\
$\Delta V_{1.1} = 2.154$	& $-1.48$ &	$-1.28$ \\
$\Delta V_{1.2} = 2.470$	& $-1.46$ &	$-1.27$ \\
$\Delta V_{1.4} = 2.950$	& $-1.44$ &	$-1.24$ \\
$S_{2.5} = 5.648$        & $-1.50$ &	$-1.25$ \\
$S_{2.0} = 6.903$	    & $-1.47$ &	$-1.26$ \\
Mean & $-1.47 \pm 0.02$ & $-1.26 \pm 0.02$ \\
\tableline 
\end{tabular}
\label{met}
\end{center}
\end{table}

\subsection{Horizontal Branch Morphology} 
\begin{figure}[t]
      \plotone{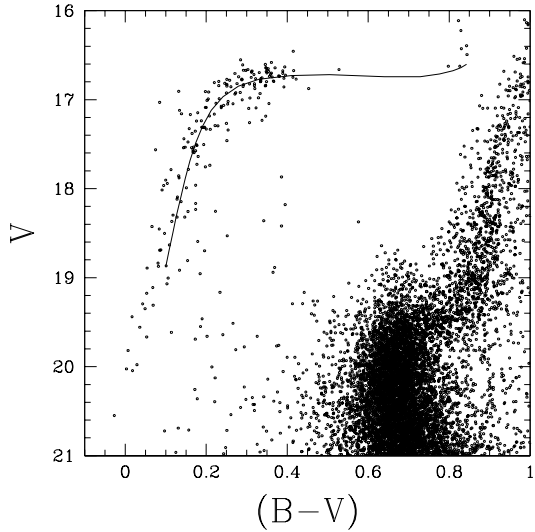}
      \caption{\footnotesize{Zoomed NGC~5286 CMD around the HB region with \citet{VBD2006} ZAHB for ${\rm [Fe/H]} = -1.71$, $[\alpha/{\rm Fe}] = 0.3$)}}
      \label{finalhb}
      \end{figure}

Figure~\ref{finalhb} shows the HB region of the \objectname{NGC~5286} CMD in the ($\bv$,{\it{V}}) plane. As before, known variable stars were omitted from this plot (see Paper II for a detailed discussion of the NGC~5286 variable star content).\\ 

We computed useful HB morphology parameters like the {\it{Lee-Zinn parameter}} {$\mathcal{L}$} (\citealp*{Z1986,Lee1990,Leeall1990}) and the {\em Buonanno parameter} ${\mathcal{P_{HB}}}$ (\citealp{Buo1993,Buo1997}),  defined as 

\begin{align}
\mathcal{L}      & \equiv  \mathcal{(B-R)/(B+V+R)},\\
\mathcal{P_{HB}} & \equiv  \mathcal{(B{\rm 2}-R)/(B+V+R)},
\end{align}

\noindent where $\mathcal{B}2$ is the number of blue HB stars bluer than ${\it{(\bv)}}_0 = -0.02$, and $\mathcal{B}$, $\mathcal{V}$, $\mathcal{R}$ are the numbers of blue, variable (RR Lyrae), and red HB stars, respectively. We also derived two parameters defined by \citet*{Pre1991}: the mean unreddened color of blue HB between $-0.02 < (\bv)_0 < 0.18$, $(\bv)_W$, and the number of HB stars in this color interval normalized by the number of HB stars with $(\bv)_0 < 0.18$, $B_W/B$. Table~\ref{hbtab} shows the HB morphology parameters for \objectname{NGC~5286}.  

\begin{table}[h!]
\begin{center}
\footnotesize
\caption{\footnotesize{HB Morphology Parameters for NGC~5286}}
\begin{tabular}{lcc}
\tableline\tableline
Parameter & Value \\
\tableline
$\mathcal{B:V:R}$	& $0.764 : 0.338 : 0.027$ \\
$\mathcal{B/(B+R)}$	& $0.973 \pm 0.03$ \\
$\mathcal{(B-R)/(B+V+R)}$	& $0.739 \pm 0.02$ \\
$\mathcal{(B{\rm 2}-R)/(B+V+R)}$	& $0.342 \pm 0.02$ \\
$(\bv)_W$	&  $0.074 \pm 0.005$ \\
$B_W/B$	&  $0.523 \pm 0.05$ \\
\tableline
\end{tabular}
\label{hbtab}
\end{center}
\end{table}

    \begin{figure}[t]
      \plotone{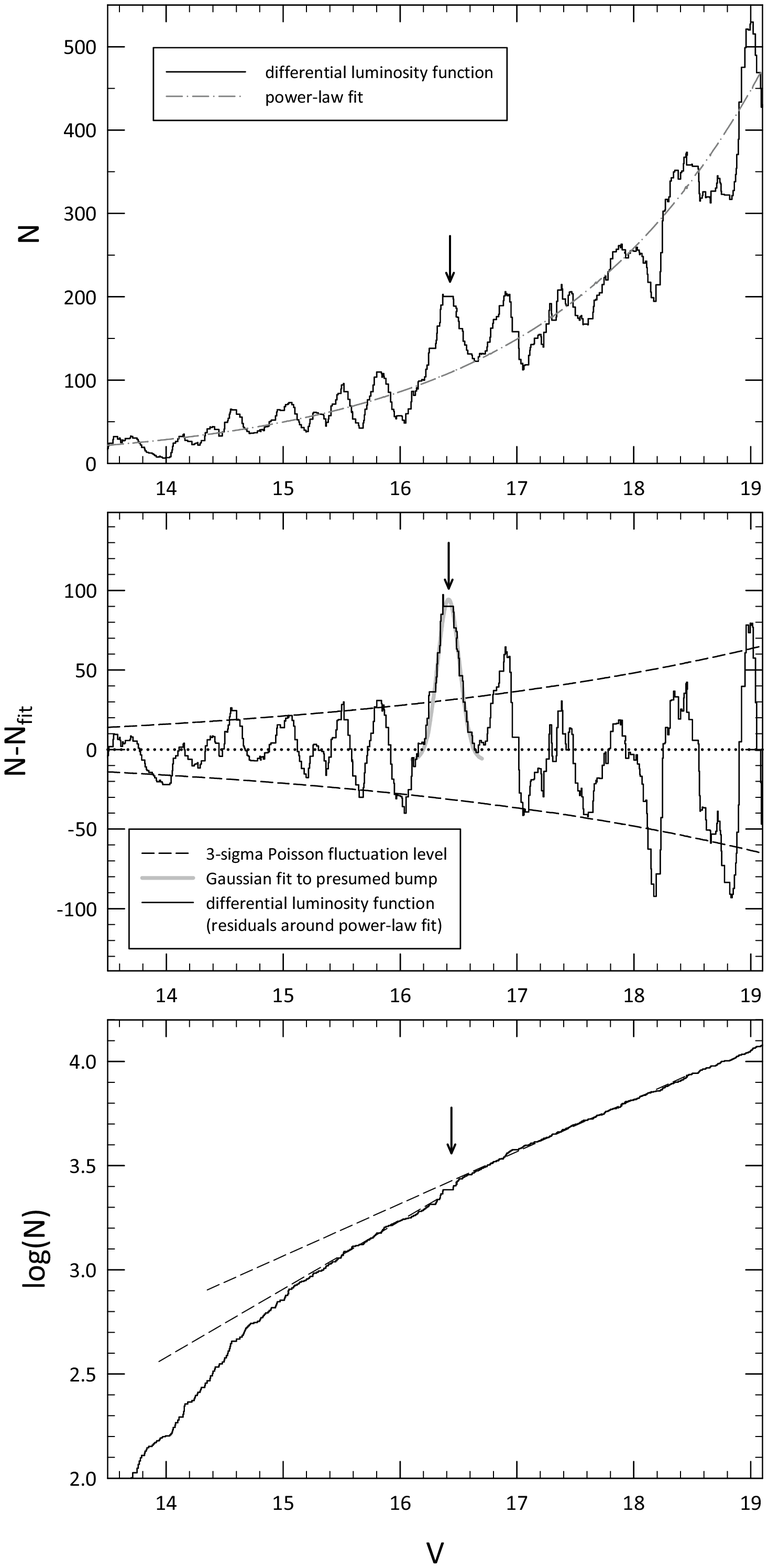}
      \caption{\footnotesize{
{\em Top}: differential RGB luminosity function for NGC~5286 ({\em solid line}). A power-law fit to the data is shown as a {\em dash-dotted gray line}. {\em Middle}: the residuals around the power-law fit are shown as a {\em solid line}. The {\em dashed lines} indicate the Poisson $3-\sigma$ level, as computed based on the derived power-law fit. The gray line shows the Gaussian fit over a $\pm 0.3$~mag region around the candidate RGB bump peak. {\em Bottom}: cumulative luminosity function, indicating the break expected if the bump is present at the indicated location. All three panels consistently indicate that the RGB bump is located at the position marked by the arrow. 
	  }}
	  \label{historgb}
      \end{figure} 
      
    \begin{figure}[t]
      \plotone{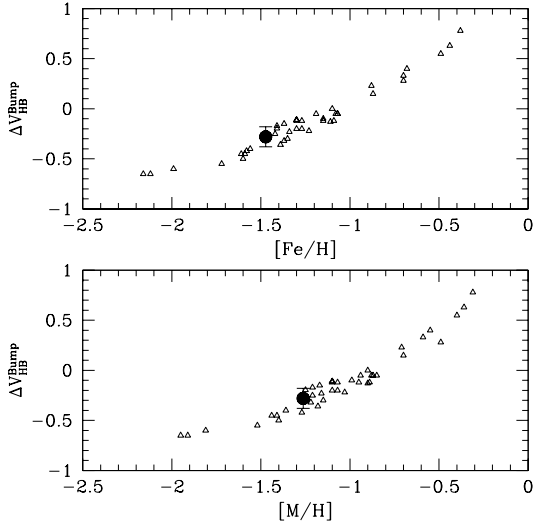}
      \caption{\footnotesize{Magnitude difference between the bump and the HB, $\Delta V^{\rm bump}_{\rm HB}$, as a function of ${\rm [Fe/H]_{CG97}}$ ({\em top}) and ${\rm [M/H]}$ ({\em bottom}). The {\em filled circle} indicates the derived position for NGC~5286, whereas the triangles correspond to data for other globular clusters, from \citet{Fer1999}. }}
      \label{f99}
      \end{figure}

	    \begin{figure}[t]
      \plotone{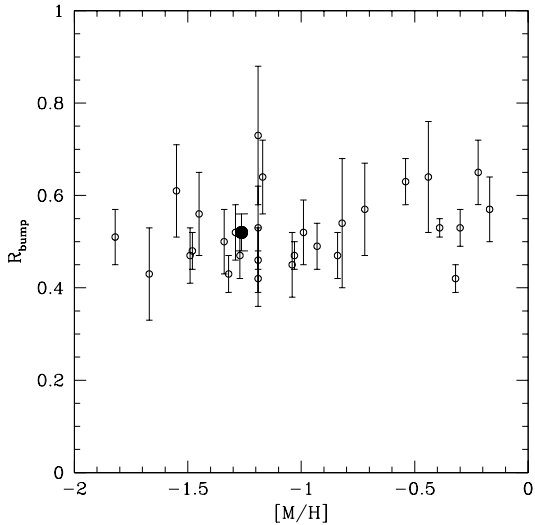}
      \caption{\footnotesize{$R_{\rm bump}$ vs global metallicity. The {\em filled circle} indicates the measured value for NGC~5286.}}
      \label{rbumpfig}
      \end{figure}

\subsection{Red Giant Branch} 
During the RGB phase, the H-burning shell advances outward in mass, leading to a continued increase in mass of the He core. Eventually, the H-burning shell actually encounters the chemical composition discontinuity that was left behind as a consequence of the maximum inward penetration of the convective envelope. Since the envelope is naturally H-rich, this means that the H-burning shell is suddenly presented with an extra supply of fuel. The structure of the star, presented with this extra fuel supply, readjusts momentarily to this new situation, with an actual (small) reversal in its direction of evolution before it resumes its ascent of the RGB. The details of this process depend crucially on the precise abundance profile in the H-burning shell (see, e.g., \citealp*{scea02}). In the observed CMDs and RGB luminosity functions (LFs) of globular star clusters, and as first predicted by \citet{hct67} and \citet{ii68}, one in fact identifies the so-called {\em RGB ``bump''} as an observed counterpart of this stellar interior phenomenon (e.g., \citealp*{ckea85,ffpea90}). Importantly, the RGB bump also appears to correspond to the position in the CMD that marks the onset of mixing of nuclearly-processed elements beyond that predicted by the canonical theory (e.g., \citealp{rgea00,dv03,cc05,rbdl07} and references therein).

To determine the position of the RGB bump in NGC~5286, we construct a smoothed RGB LF, following the procedure described in \citet*{zoc99}. This already reveals a candidate bump at $V \simeq 16.4$ (Fig.~\ref{historgb}, {\em upper panel}. To confirm that this is a significant detection, we carry out a power-law fit to this LF, which is shown as a dash-dotted gray line in Figure~\ref{historgb}. We then calculate the difference between the actual smoothed RGB LF and the power-law fit, and compare the result with the expected $3-\sigma$ Poisson fluctuation level. As shown in Figure~\ref{historgb} ({\em middle panel}), the most significant feature in the RGB LF does indeed lie at the position previously suspected to correspond to the RGB bump. We carry out a Gaussian fit to the data within a $\pm 0.3$~mag range around the detected feature, which leads to the curve shown as a gray line in Figure~\ref{historgb} ({\em middle panel}). The parameters of the fit are as follows: $V_{\rm bump} = 16.415 \pm 0.002$~mag, $\sigma_{\rm bump} = 0.093 \pm 0.004$~mag. We confirm this detection using the integrated RGB LF method of \citet{ffpea90} (Figure~\ref{historgb}, {\em bottom panel}), and finally adopt for the position of the bump a value $V_{\rm bump} = 16.41 \pm 0.05$~mag~-- where the adopted error bar represents a compromise between the width of the Gaussian fit and the formal error in the position of its center. 

We next compute the difference in brightness between the RGB bump and the ZAHB level 
(defined as in \S\ref{sec:metred}), obtaining $\Delta V^{\rm bump}_{\rm HB} = -0.28 \pm 0.05$. A plot showing $\Delta V^{\rm bump}_{\rm HB}$ as a function of metallicity is presented in Figure~\ref{f99}. The cluster data are from Table~5 in F99. In this figure, the derived position of NGC~5286 is shown as a filled dot with error bar; as we can see, the position of NGC~5286 in this plane is consistent with our inferred metallicity for the cluster. 

We also calculated $R_{\rm bump}$ \citep{Betal2001}, the ratio between the number of RGB stars in the bump region ($V_{\rm bump} \pm 0.4$) and the number of RGB stars in the interval $V_{\rm bump} + 0.5 < V < V_{\rm bump} + 1.5$. This quantity is important in revealing whether deep mixing can start prior to the bump level or not. In our case we obtained $R_{\rm bump} = 0.52 \pm 0.04$. Comparision of the measured $R_{\rm bump}$ parameter with the cluster sample of \citealp{Betal2001} (their Table~1) shows good agreement, as can be seen from Figure~\ref{rbumpfig}. This suggests, as discussed by \citeauthor{Betal2001}, that an important amount of deep mixing does not take place before stars reach the bump.

\section{Comparison with M3} \label{sec:vsM3}
\begin{figure}[t]
  \plotone{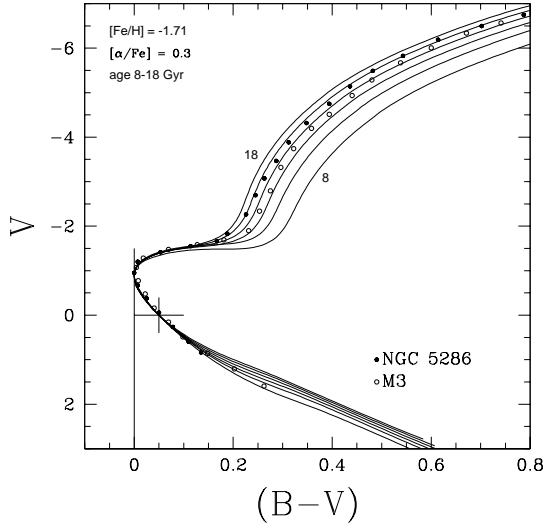}
  \caption{\footnotesize{Comparision of the M3 ({\em crosses}) and NGC~5286 ({\em points}) ridgelines with theoretical isochrones \citep[from][]{VBD2006} for ${\rm [Fe/H]} = -1.71$, $[\alpha/{\rm Fe}] = 0.3$, and ages ranging from 8 to 18~Gyr, in steps of 2~Gyr ({\em lines}). The models and data were registered as recommended in \citet{stetal1999}. }}
  \label{isoc}
  \end{figure}

We compare \objectname{NGC~5286} with \objectname{M3}, a well studied cluster of similar metallicity 
\citep[${\rm [Fe/H]} = -1.57$, according to][]{harr1996}. Figure~\ref{isoc} shows the fiducial points for \objectname{M3}, as derived by us following the same procedure as in \S\ref{sec:CMD} for NGC~5286, compared with those for \objectname{NGC~5286}, in the ($\bv$, $V$) plane. The M3 photometry 
was kindly provided by P. B. Stetson (2008, priv. comm.; see also \citeauthor{psea05} \citeyear{psea05}). \citet{VBD2006} isochrones for ${\rm [Fe/H]} = -1.71$, $[\alpha/{\rm Fe}] = 0.3$, and ages ranging from 8 to 18~Gyr are also shown. We chose the metallicity nearest to the one that we found for \objectname{NGC~5286} in the ZW84 scale (see \S\ref{sec:metred}). A variation in the metallicity at the level of the difference between the results in the ZW84 and CG97 is not expected to affect substantially the estimation of relative ages (see, e.g., \citealp*{VBS1990}).

As in \citet{stetal1999}, the isochrones and fiducial points for both clusters were shifted horizontally to match each other's turnoff colors ({\em vertical line}), and then shifted vertically to register the point on the upper MS that lies 0.05~mag redder than the turnoff ({\em cross}). According to this figure, and on the basis of equation~(1) in \citet{vs90}, assuming an identical ${\rm [O/Fe]} = 0.3$ for both clusters, \objectname{NGC~5286} seems to be around $1.7 \pm 0.9$~Gyr older than \objectname{M3}~-- which goes in the right sense to explain the former's bluer HB.

As well known, another method to derive GC ages is the so-called {\em $\Delta V$ method}, which is based on the difference in magnitude between the HB and the TO levels. In this case we follow the procedure described by \citet*{chab96}, using their equation~(2) and assuming $M_V({\rm RR}) = 0.20 \, {\rm [Fe/H]} + 0.98$. We decided to use this specific calibration for $M_V({\rm RR})$ because is gives the closest match to the absolute magnitudes of RR Lyrae stars recently calibrated by \citet{cacor08}. To obtain the position of the TO and HB levels for M3, we use the same procedure as for NGC~5286. Then we used $V({\rm HB}) = V_{{\rm ZAHB}} - 0.05 \, {\rm [Fe/H]} - 0.2$, again as in \citeauthor{chab96}. We thus obtained $\Delta V^{TO}_{HB} = 3.48\pm 0.1$ for NGC~5286, and $\Delta V^{TO}_{HB} = 3.39 \pm 0.1$ for M3. These values imply that \objectname{NGC~5286} is around $1.5 \pm 1.6$~Gyr older than \objectname{M3}, in good agreement with the value previously derived on the basis of the horizontal method. 

We refrain from providing an absolute age for NGC~5286 in this paper, given the well known problems in accurately defining a proper age scale for Galactic globular clusters, but do note that, in case the above results are correct, NGC~5286 may rank among the oldest globulars in our galaxy \citep[compare, for instance, with Table~4 in][]{dv00}. This also implies that, if NGC~5286 is a bona fide member of the Canis Major dSph galaxy, then the latter's oldest components may be at least as old as the oldest Milky Way GCs. 

\section{Summary} \label{sec:summ}
We have obtained {\it{BV}} CCD photometry for the globular cluster \objectname{NGC~5286}, extending about two magnitudes deeper than the MS turnoff. The latter is found to be aproximately at $V_{\rm TO} = 20.05 \pm 0.1$ mag and $(\bv) = 0.66 \pm 0.02$. We detect the RGB bump at $V_{\rm bump} = 16.41 \pm 0.05$~mag.

 A variety of HB morphology parameters was also computed. A metallicity of ${\rm [Fe/H]} = -1.70 \pm 0.10$ was derived in the ZW84 scale based on several RGB parameters, in good agreement with \citet{harr1996}. 

A comparision between the \objectname{NGC~5286} and \objectname{M3} CMDs with theoretical isochrones for ${\rm [Fe/H]} = -1.71$ and $[\alpha/{\rm Fe}] = 0.3$ suggests that \objectname{NGC~5286} is around $1.7 \pm 0.9$~Gyr older than \objectname{M3}, which goes in the right sense to explain the former's bluer HB morphology. If NGC~5286 is indeed a bona fide member of the Canis Major dSph galaxy, this implies that the latter's oldest components may be at least as old as the oldest Milky Way GCs.

\acknowledgments
We thank the referee for several constructive comments that led to a significant improvement in the presentation of our results. M. Zorotovic and MC acknowledge support by Proyecto Fondecyt Regular \#1071002, M. Zoccali acknowledges support by Proyecto Fondecyt Regular \#1085278. HAS was supported by CSCE and NSF grant AST 0607249.

\end{document}